# The Australian PCEHR system: Ensuring Privacy and Security through an Improved Access Control Mechanism

P. Vimalachandran [1,*], H. Wang [2], Y. Zhang [3] and G. Zhuo [4]

[1] Centre for Applied Informatics, College of Engineering and Science, Victoria University, Melbourne
[2,3] Centre for Applied Informatics, College of Engineering and Science, Victoria University, Melbourne
[4] Department of Computer Science, Taiyuan Normal University, China

## Abstract

An Electronic Health Record (EHR) is designed to store diverse data accurately from a range of health care providers and to capture the status of a patient by a range of health care providers across time. Realising the numerous benefits of the system, EHR adoption is growing globally and many countries invest heavily in electronic health systems. In Australia, the Government invested $467 million to build key components of the Personally Controlled Electronic Health Record (PCEHR) system in July 2012. However, in the last three years, the uptake from individuals and health care providers has not been satisfactory. Unauthorised access of the PCEHR was one of the major barriers. We propose an improved access control model for the PCEHR system to resolve the unauthorised access issue. We discuss the unauthorised access issue with real examples and present a potential solution to overcome the issue to make the PCEHR system a success in Australia.

*Corresponding author. Email: Pasupathy.Vimalachandran@live.vu.edu.au

## 1. Introduction

During the last two decades, modern technology has increasingly being used in the healthcare sector in order to enhance the quality and the cost efficiency of the healthcare services. EHR is one of them. In most parts of the developed world, healthcare has evolved to a point where patients have more than one healthcare provider. This has resulted in the growing need to create an integrated infrastructure for the collection of diverse medical data for healthcare professionals, where the adoption of standardised Electronic Health Record (EHR) has become imminent. An EHR is a summary of health events usually drawn from several electronic medical records and may consist of the elements that are eventually shared in a national EHR [1], [2]. An online EHR enables patients to manage and contribute to their own medical notes in a centralised way which greatly facilitates the storage, access and sharing of personal health data. It is clear that storing medical records digitally on the cloud offers great promise for increasing the efficiency of the healthcare system.

### 1.1. Personally Controlled Electronic Health Record (PCEHR)

A national EHR was introduced to Australia in 2012 and the Government has invested multi millions of dollars to build key components of the Personally Controlled

Electronic Health Record (PCEHR) to improve health outcomes and reduce costs for health in the country [3].

However the take-up by individuals (patients or consumer) and health care providers of the PCEHR system has been inadequate. The government has failed to meet a self-set target of 500,000 registrations of its PCEHR by July 1, 2013 [4], doctors have uploaded just 41,998 shared health summaries onto these records (more than 2 million e-health records are empty) and the scheme has so far cost taxpayers more than $ 1 billion to develop, or almost $24,000 per shared health summary [5].

The questions still remain, why the PCEHR has not met its targets and the take-up did not reach the expectations. First of all, setting up an EHR system faces many challenges which ultimately impede its wider adoption. A privacy and confidentiality concern is one of the top ones. Addressing these concerns to win individuals mindset is crucial. Once patients' personal health data are stored in the cloud or local server with PCEHR, it is not quite clear who else can access it other than the patient's usual doctor. For example, with the current system, in a health care provider organisation, all other health care providers working for the organisation can access patient clinical information. There are also instances' where administration staff may access patients' clinical information for improving the business (e.g. targeting chronic disease high risk or pap smear patients, who are due for a reminder) [6]. Externally it is also not sure who can access those data once it is stored. Department of Health (DOH) in [7] points out that there are very limited circumstances where anyone, including the Government, may access someone's PCEHR documents. This statement makes more uncertainty for individuals.

Consequently, consumers want evidence that their personal health information is protected and remain confidential when stored on the PCEHR. Consumers have argued that 'the best way to protect privacy is for consumers to have ultimate control over who has access to their record [8].

Consumers also find difficulties in accessing their health information from EHR or electronic medical records. Sometimes, they must pay to access their own health information. This is frustrating, and the current PCEHR system must resolve this issue and give free access to their own health information. Overwhelmingly, consumers want to have access to their own records [8]. However the current PCEHR system did not reach that far yet. However, on the other hand, consumer should not be given permission to modify or remove uploaded documents for themselves. Once the document is uploaded for a patient whether it is Shared Health Summary or Event Summary (ES), the patient can only view (read only mode) the details. Patients cannot modify any details in it because when health care providers create records it is turned to Clinical Document Architecture (CDA) type which is non-changeable format like Portable Document Format (PDF).

With the current arrangement of the PCEHR, the patients cannot modify the information that is in their record, however they could delete the whole record [9] as shown in Figure 1 below. This is again an arguable matter for the health care providers. Once the health care provider and the patient decide the information is valid and useful for future treatment and then uploaded, why then does the patient have the option to delete the whole SHS or ES from their record?. Health care providers, therefore argue, how can they depend on the records when options available for patients to remove the records.

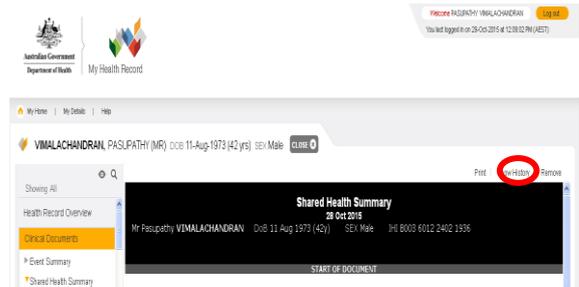

Figure 1: Option for a Patient

On the other hand, the Department of Health (DOH) justifies [8] that it is patients' record; hence the patients are able to remove a document or information about prescribed medications or other treatment from the record.

As mentioned in Figure 2, the patients also have control to restrict some health care provider organisations from viewing any or all of the record(s) from the PCEHR. This is again questionable for health care providers.

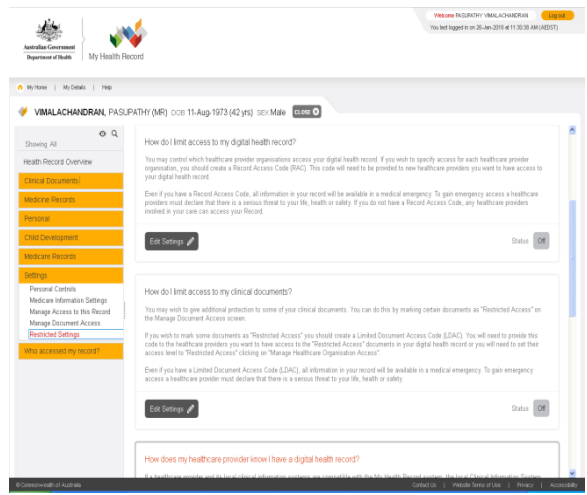

Figure 2: Patient Option to give Restrict Access

The DOH explains [10] this control gives option for patients who can access their record. However health care providers worry that the useful health information will not be accessible for the effective treatment, even in an emergency if they block something relevant.

This ability of the patient to hide aspects of their record is, finally, not only incomplete and uncertain of the health information integrity, but it becomes a significant clinical risk. David Glance [11] mentioned that the first problem is the system still represents a "scrapbook" approach to a clinical record. There is no guarantee that all the health professionals involved in the care of a patient will participate and supply information, or that the information supplied will be complete.

Australian Medical Association [12] details that the personal record is only a "memory prompt" for the patient and that "remains the treating medical practitioner's responsibility to take a clinical history from their patient". This means again, a doctor cannot rely on the health record to make clinical decision and the benefit of the system would be limited. A SHS or ES would be beneficial if the record is a fully distributed and shareable clinical record that all health care providers involved in the care of the patient would have access to the record or could appropriate access when necessary.

It is important to acknowledge and investigate these challenges and shortcomings associated with the current electronic health information system and to determine possible solutions to ensure its wide adoption and the success of the PCEHR system in Australia.

## 1.2. Access Control of the PCEHR

There is significant concern about the privacy of data on PCEHR and its potential misuse [13]. In healthcare organisations, there are non-clinical staff, such as administration staff who may need to access clinical related information including PCEHR to target patients to increase the organisation's business. For example, the practice follows up with health checks due and reminds mainstream patients or identified chronic disease high risk patients of the need for consultations. In these circumstances, administration staff may access clinical information. This access may lead to internal abuse. Therefore, administration staff accessing clinical information is a high risk.

Furthermore the system operator of the PCEHR who manages the system may intentionally leak patients' clinical information. The access control currently in use does not prevent this kind of breach.

In computer security, access control covers authentication, authorisation and audit. Access control systems provide the important services of identification, authentication, authorisation and accountability to enter into an application or system. Identification and authentication determine who can log into a system (the system may be an application or even an operating system). Authorisation provides different privileges for a system (usually categorised high-level, medium-level and low-level) in accordance with the employee's role in a health care organisation. Finally the accountability identifies the subject a user worked on during his or her log-in.

## 1.3. Our Contribution

We review and understand previous attacks and methods utilised to obtain sensitive health information from databases. This will assist to implement an improved access control disclosure method in the future (Section 2). We discuss access control problem with real world examples when using the PCEHR systems in a health care provider organisation environment (Section 3). We propose an appropriate access control mechanism that could be used in Australian health care settings to improve privacy and security. This model will prevent unauthorised access and misuse within health care provider organisations. Initial development of the proposed model is explained with a sample computer programming language to allow a super user to give permission for usual users to access sensitive health information in day to day activities. A Mobile Security System (MSS) is also suggested for the communication between super users and users to ensure prompt uninterrupted permission (Section 4). Finally, Section 5 concludes the paper with future research directions.

## 2. Related Work

There are different access control strategies for Electronic Medical Record (EMR) and EHR that have been developed in the past [14].

According to one Forrester study, 80% of data security and privacy breaches involve insiders, employees or those with internal access to an organisation, putting information at risk [15]. With health sensitive data, this risk becomes more prominent. Many researchers have proposed various resolutions to solve the security and privacy problems associated with the EMRs and EHRs. These problems mainly refer to access control. The term "access control" is simply defined as "the ability to permit or deny the use of something by someone" [15]. The key objective of access control mechanisms is to permit authorised users to manipulate data and thus maintain the privacy of data [16]. There are different access control mechanisms that have been identified in the literature review. The basic models of the access control principles are i) Discretionary Access Control (DAC), ii) Mandatory Access Control (MAC), iii) Role Based Access Control (RBAC) and iv) Purpose Based Access Control (PBAC). However, the development is not satisfactory enough to fulfil the privacy requirements of EMRs and EHRs [17].

DAC uses access restriction set by the owner and restricts access to the objects. However a user who is allowed to access an object by the owner of the object has the capability to pass on the access right to other users without the involvement of the owner of the object [18]. Because of this granting, read access transitive, the policies are open for Trojan Horse Attack [19].

MAC is a set of security and privacy policies constrained according to system classification, configuration and authentication. The policies made by a central authority [20]. Compared to DAC, MAC policy can prevent a Trojan Horse Attack and the integrity of the data objects can be protected by using the "Read Up" and "Write Down" Rules. In MAC, the individual owner of an object has no right to control the access. Therefore, MAC policy fails to preserve the privacy requirement for EHRs of the patients [21]

In RBAC [19], each user's access right is determined based on user roles and the role-specific privileges associated with them. RBAC policy uses the need-to-know principle to assign permissions to roles and to fulfil the least privileged condition by the system administrator. However, RBAC does not integrate other access parameters or related data that are significant in allowing access to the user [22]. PBAC is based on the notion of relating data objects with purposes [23]. Many researchers have identified that greater privacy preservation is possible by assigning objects with purposes [24]. However, Al-Fedaghi describes [25] that PBAC leads to a great deal of complexity at the access control level.

In addition to access control mechanisms, it is also important to identify the spectrum of attacks or misuse that could be performed by attackers. A wide range of attacks have been documented in the literature. It is essential to know the different possible attacks for health based databases, in order to design a suitable health data security system. To achieve this goal the literature review has been performed to discuss different main attacks that health based databases currently face.

In the British Computer Society website at http://www.bcs.org/server.php?show=ConWebDoc.8852, Amichai Schulman and Imperva say "enterprise database infrastructures, which often contain the crown jewels of an organisation, are subject to a wide range of attacks" [26].

The reviewing and understanding of previous attacks will assist to prevent the access control disclosure in the future. Following main methods utilised to obtain sensitive health information from databases.

1. Excessive privilege granted to staff
2. Privilege abuse
3. Unauthorised privilege elevation
4. Platform vulnerabilities
5. SQL injection
6. Weak audit
7. Weak authentication
8. Exposure of back-up data

With excessive privilege, healthcare organisation application users are granted privileges that may exceed the requirements of their role. As an example, a reception/ administrative staff member whose job requires name, contact details and time of the appointments of a patient, may be able to view clinical notes of patients.

Healthcare application users may abuse legitimate data access privileges for unauthorised purposes. This is known as 'privilege abuse'.

Unauthorised privilege elevation means that the attackers may take advantage of vulnerabilities in health based cloud software systems to convert low-level access privileges to high-level access privileges. For instance, an attacker may take advantage of cloud based system buffer overflow vulnerability to grant administrative privileges.

Platform vulnerability is taking advantage of the vulnerabilities in underlying operating systems, which may lead to unauthorised data access or corruption. The blaster worm took advantage of a Windows 2000 vulnerability to take down target servers [27].

Users may take advantage of vulnerabilities in front-end web applications and stored procedures to send unauthorised database queries. This is known as "SQL injection".

Weak audit policy and technology represents risks in terms of compliance, deterrence, detection, forensics and recovery. In other words, the cloud based health system software provides weak audit solutions itself. These products very rarely log the detail about what application was used; the source IP address and what queries failed.

Weak authentication allows attackers to assume the identity of legitimate database users. Most of the time, the users use their name, personal identification, meaningful words or plain text as a password.

In most situations, people protect the main cloud based health database, not actual back-ups. With exposure of back-up data, attacks have involved theft of database backup tapes and hard disks.

# 3. Access Control and PCEHR

A good information security must have following three main characteristics [28], [29]:

- Confidentiality – the prevention of unauthorised disclosure of the information.
- Integrity – the prevention of unauthorised modification of the information.
- Availability – the prevention of unauthorised withholding of the information.

Privacy is the right of an individual to not have their private information exposed, whilst confidentiality is limiting access to information to authorised individuals only. Confidentiality is often used interchangeably with privacy but they are not exactly the same.

For an EHR, as a secure information system which holds very sensitive and useful health information, obligating the above characteristics is paramount. These characteristics must ensure security and smooth operations of the following three stages of an EHR:

- to add health information for a patient by all involved health care providers.
- to protect the information holding of a patient while storing.

- to ensure availability of the information when required by health care providers.

Maintaining above three stages for an EHR is a big challenge. The system must make it easy for an authorised user to gain access to the information whilst preventing unauthorised access.

In order to securely access information within information systems identification (e.g. username), authentication (e.g. password) and authorisation (e.g. access rights) are required. Access control is conceptually part of the authorisation process that checks if a user can access the information that requested.

Health care provider organisations use the PCEHR through their clinical software systems. The software systems use DAC and MAC access control principles. RBAC is also used in those systems however PBAC is not in use in many healthcare systems because of the complexity at the access level. Furthermore, considering the current privacy and security issues associated with health records, a single access control principle is inadequate to protect the highly sensitive information. Thus, it is crucial to use a combination of more than one access control principle in this environment. When administration staff access clinical information from the system where RBAC is switched on, the purpose of the access is not mentioned. To solve this issue, an authorisation from an authority must be given to access the information. Then a combination of RBAC and PBAC must be applied for a secure access. This means, if an administration staff wants to access the clinical information, a high-level management staff must give permission every time. The high-level management staff might be a doctor or a nurse or practice manager who has high-level privilege to access all components of the health record. This requires both access control principles RBAC and PBAC to access the proposed model and it will satisfy the requirement.

In a healthcare provider organisation or an organisation that manages the PCEHR system, it is not clear who accesses what information within the organisation. In a general practice (medium or large) environment in Australia, organisations normally use two types of software systems to deal with patients. One is the Patient Management System (PMS) that assists with appointment and billing related activities. This is also known as 'billing system'. The other is for managing clinical associated activities and recording medical information which is called 'clinical system'. Most general practice software systems are integrated with both systems clinical and billing. In some cases, the same product has two different software systems that are compatible and work together to manage both clinical and billing activities. If an organisation uses different software systems for billing and clinical, then assigning access control is easier. For instance, reception staff has access to the billing system and not to the clinical system. On the other hand, clinicians including doctors and nurses access both clinical and PMS but not billing. However, if an organisation uses an integrated one system for both billing and clinical, then the issues associated with access control becomes complicated. However there are situations, where healthcare organisations manage this issue by giving permission levels based on the roles and purposes. The software itself manages these permission controls.

## 4. Proposed Model and Development

In health care organisations current settings, considering the financial benefit to the organisation, non-clinical staff such as administration staff accessing clinical information including PCEHR documents cannot just be ignored. Thus, on the other hand, it is a high risk when the staff accessing the PCEHR can intentionally leak patients' clinical information. The access control currently in use does not prevent this kind of breach. There should be, consequently, a strong access control mechanism in place, which promises to prevent unauthorised access of the PCEHR. We strongly believe our new concept "Log-in-Pair" would be an ideal answer to protect or minimise the unauthorised disclosure. The development of the "Log-in-Pair" model is also explained with a sample coding in this section below.

To prevent unauthorised access and/ or disclosure to the PCEHR system, the "Log-in-Pair" could be used. In this model, to access health information, an employee who has top level privilege (called "super user") has to give authorisation to a user to access the health sensitive data. Hence, the super user keeps track of what the user does with the sensitive data. The users are well-known that the super-users have given permission to access the sensitive information and they can keep track of what is being accessed. It is like a counter check. The responsibility and the accountability are shared.

The "Log-in-Pair" proposed model covers both access control principles RBAC and PBAC to access sensitive clinical information including PCEHR documents. Thus, the proposed model will not only ensure a higher level of security but will also resolve the unauthorised access problem that previously discussed in Section 2.

In this setting, every user has their own individual user identity and password (alternatively the same credentials that user uses for the clinical system can be used) to enter into the system. In this pair log-in concept, see Table 1 below, for user A to enter into the system the super user D or E or F will receive an alert to give permission. The permission may simply be a touch on their mobile phone using MSS. In this example, a healthcare organisation with three users (A, B and C) and three super-users (D, E and F) is considered.

Table 1: Basic pair design

| Pair | Users & Super Users |
|------|---------------------|
| 1 | A & D or E or F |
| 2 | B & D or E or F |
| 3 | C & D or E or F |

The log-in page is designed to accept input from a user and then to send an alert to the assigned super-users. Assigning more than one super-user will save time and give more certainty for a user. The security assurance in this system is that one person cannot function on their own to access the sensitive information from the PCEHR. If a user needs to access the system, they must obtain a super-user permission. This arrangement must prevent unauthorised access and abuse of the PCEHR.

This system has its own problems. The following problems have been identified in the proposed model. Table 2 below illustrates the identified problems with the proposed system and potential solutions for those problems.

Table 2: Identified problems and the solutions of the proposed model

| Problem | Solution |
|---------|----------|
| If all super users are absent (D, E and F), a user cannot access the system and complete the routine activities. | MSS helps to send alerts to super users' mobile devices. Even though they are on leave it is possible for them to give permission through the mobile devices e.g. mobile phone. Assigning more than one super-user for a user also will increase the accessibility. |
| The system cannot prevent if both user and super-user choose to abuse the sensitive data. | System Monitoring Facility (SMF) that suggested as future development in Section 5 should detect these sorts of abuses. |
| Accessing more than one user at a time could create potential bottleneck issues of the system. | The system itself must be notified and does not give access to other users to avoid bottlenecks and unnecessary delays in logging on the system e.g. auto tracking |
| If doctors and nurses are potential "gatekeepers" (super-users), these professions are already extremely busy, and likely to create users circumventing the system. | The super-users can simply give the permission by one touch on their mobile screen from anywhere. |

Every non-clinical user who need to access clinical information and the PCEHR must be assigned to a super-user. The source code to create a super-user for each user is illustrated in Code 1.

```
Private Sub cmdok_Click()
bcheck = checkdata
usertype = Left(cmbUsertype.Text, 1)
 If bcheck = checkdata Then
rs.Open "select * from usertable where userid='" +
txtuser.Text + "'", cn
        If Not rs.EOF And Not rs.BOFThen
MsgBox "This user already exists", , "HighSec System"
        Else
newpwd = encryptdata(txtpassword.Text, newkey)
newpwd = txtpassword.Text
ssql = "insert into usertable (userid,pwd,usertype)
values('" + txtuser.Text + "','" + newpwd + "','" +
usertype + "')"
InputBox "", ,ssql

cn.Executessql
        If usertype = "N" Then
ssql = "insert into groupuser (user1,user2) values('" +
txtuser.Text + "','" + cmbManager.Text + "')"
cn.Executessql
        End If
ans = MsgBox("User created succesfully. " + vbCrLf + "
Do you want to close this window?", vbYesNo)
        If ans = vbYes Then
        Unload Me
        Else
txtuser.Text = ""
txtpassword.Text = ""
        End If
    End If
rs.Close
    End If
End Sub

Private Sub Form_Load()
ssql = "select * from usertable where usertype='M'"
rs.Openssql, cn
While Notrs.EOF
cmbManager.AddItemrs(0)
rs.MoveNext
Wend
rs.Close
cmbUsertype.ListIndex = 0
End Sub
```

Code 1: Creating super-user and the verification process

The verification process that the code 1 includes will check that the user and the super-user pair are correct before the permission request has been sent.

A MSS would be a best solution for this new concept. Once the mobile security two-factor authentication is enrolled, the user logs into the system using their usual username and password. Then the super-user receives a message through the device for the permission. The permission could be given via Short Message Service (SMS), voice call, one-time pass code or a mobile smart phone apps.

For example, 'Duo' MSS [30] has its own smart phone app to do the two-factor authentication verification process. The system also lets super-user link multiple devices to the account such as mobile phone and a landline, a landline and hardware token or two different mobile devices [30]. This will provide increased accessibility for super-users and prompt and more certain access for users. As mentioned in Figure 3 below, the MSS will provide an additional security layer for the PCEHR system. The username and password that has been created for the system remain the same and once provided, the super-user approval request will be sent out to the super-user preferred communication device/s. Consequently, if a user (non-clinical staff in a health care provider organisation) wants to access patients' sensitive health information for any reason (e.g. targeting high chronic disease patients to send a reminder), then an authorised person (super-user) must give permission.

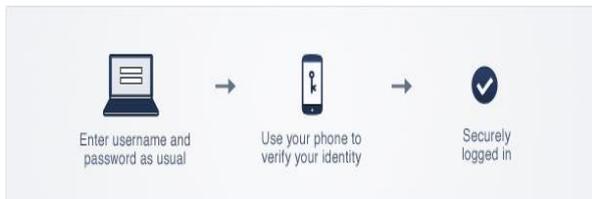

Figure 3: How a mobile security system works (source: [30])

When creating user login for a user to access health care providers' clinical software, a super-user's connection must be established as illustrated through the above computer program code (Code 1)

Moreover a super-user has more than one option to approve or deny the user login request. For example, the following options are available with Duo mobile security system.

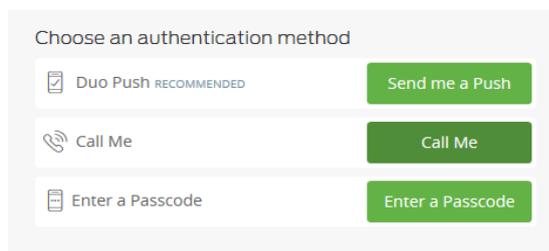

Figure 4: Authentication methods

Figure 4 above illustrates three various options that are available for super-users to give permission for users by Duo MSS.

# 5. Conclusion

In this paper, we have introduced a new concept called 'Log-in-Pair' to ensure privacy and security of the PCEHR system. Improving privacy and security of the system will guarantee the success of the system in Australia. We believe the 'Log-in-Pair' concept will be the ideal answer to minimise misuse and/or abuse within our healthcare organisations. Even though access control is the first and basic security level for any computer system, it is important to make sure that the level of protection is strong. Although the proposed method seems easy to implement, in practice, there will be more concerns when this concept is in use. However, we feel this concept could be implemented through education of policies and procedures. Ultimately, it would be one of the better solutions in maintaining and practicing a high security system in a healthcare environment. Mobile security system will ensure efficiency of the communication between users and super-users of the access control level. The super-users such as doctors, practice managers and nurses are extremely busy and are often unable to view computer screen continuously to give permission when needed. Thus, mobile security system can be communicated via mobile phones (alerts / SMSs) or other mobile devices to give the requested permission. The mobile security system also increases efficiency and effectiveness of the proposed model. A System Monitor Facility (SMF) would be beneficial to monitor users' activities within the PCEHR system and to maintain audit controls. The SMF will increase the level of security for access control as well. Therefore, developing a SMF system and creating an appropriate policy and procedure document to maintain the design of the 'Log-in-Pair' and to monitor the SMF need to be considered for future development.